\documentstyle[12pt,cite,epsfig]{article}

\renewcommand{\today}{4 April, 1996}

\newcommand{\nc}{\newcommand}
\nc{\be}{\begin{equation}}
\nc{\ee}{\end{equation}}
\nc{\bea}{\begin{eqnarray}}
\nc{\eea}{\end{eqnarray}}
\nc{\beas}{\begin{eqnarray*}}
\nc{\eeas}{\end{eqnarray*}}
\nc{\noi}{\noindent}
\nc{\sD}{\not \! \! D}
\nc{\s}[1]{\not \! #1}
\nc{\non}{\nonumber}
\nc{\bb}{\bibitem}
\nc{\lf}{\left}
\nc{\r}{\right}
\nc{\mb}[1]{\makebox[#1]{}}
\nc{\pa}{\partial}
\nc{\sA}{\not \! \! A}
\nc{\newsec}[1]{\section{#1}\mb{0.5cm}}
\nc{\h}{\frac{1}{2}}
\nc{\ra}{\rightarrow}
\nc{\la}{\leftarrow}
\nc{\ep}{$e^+e^-\ra\pi^+\pi^-\;$}
\nc{\epp}{$e^+e^-\ra\pi^+\pi^0\pi^-\;$}
\def\mathunderaccent#1{\let\theaccent#1\mathpalette\putaccentunder}
\def\putaccentunder#1#2{\oalign{$#1#2$\crcr\hidewidth
\vbox to.2ex{\hbox{$#1\theaccent{}$}\vss}\hidewidth}}

\nc{\ti}{\mathunderaccent\tilde}
\nc{\M}{{\cal M}}
\nc{\rw}{$\rho\!-\!\omega\;$}

\begin{document}
\thispagestyle{empty}
\begin{flushright}
ADP-96-13/T216 \\
hep-ph/9604375

\end{flushright}

\begin{center}
{\large{\bf Recent developments in
 rho-omega mixing}}\\
 
 {\large{\bf [Aust. J. Phys. 50 (1997) 255] }} 
 
\vspace{2.2 cm}
Heath B. O'Connell\\
\vspace{1.2 cm}
{\it
Department of Physics and Mathematical Physics \\
University of Adelaide 5005, Australia } \\
\vspace{1.2 cm}
\today
\vspace{1.2 cm}
\begin{abstract}

The topic of \rw mixing has received renewed interest in recent years and has
been studied using a variety of modern techniques.  A brief history of the
subject is presented before summarising recent developments in the field. The
present status of our understanding is discussed.

\end{abstract}
\end{center}
\vfill
\begin{flushleft}
E-mail: hoconnel@physics.adelaide.edu.au \\
{\it Refereed version of
talk given at ``Quarks, Hadrons and Nuclei" Joint Australian-Japanese
meeting at the Institute for Theoretical Physics, Adelaide, November 1995.
}
\end{flushleft}

\newpage

\begin{figure}[htb]
  \centering{\
     \epsfig{angle=0,figure=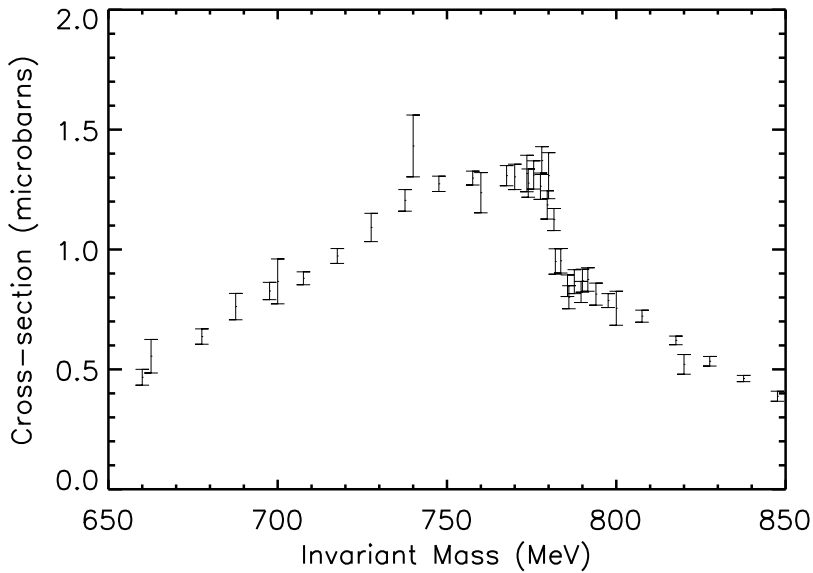,height=8.5cm}
               }
\parbox{130mm}{\caption
{The cross-section for the reaction \protect{\ep} from the data of
Ref. \protect{\cite{data}} in the \protect{\rw} resonance region. }
\label{cross}}
\end{figure}

No discussion of \rw mixing can be self-contained without a brief mention of
the Vector Meson Dominance (VMD) model, within which it is studied
\cite{Sak,review}.  A simple example to introduce VMD is the 
electromagnetic (EM) pion
form-factor, $F_\pi(q^2)$. This quantity, represents the multiplicative
deviation from the amplitude for the reaction \ep 
for the coupling of photons
to purely point-like pions.  A resonance peak is observed in the
cross-section (see Fig.~\ref{cross} showing data from
\cite{data}). 
While this
arises from low-energy, non-perturbative QCD
processes \cite{roberts}, it can be modelled by assuming the photon couples
to pions via vector mesons, the dominant one here being the rho-meson. VMD
assumes the photon interacts with hadrons through vector mesons and
mathematically, the enhancement seen in the cross-section is provided by the
pole in the meson propagator at the $\rho$ mass-point. The broadness of the peak
seen in the reaction is explained by attributing a ``width," $\Gamma_\rho$, to
the rho-meson corresponding to shifting the pole off the real axis, so that it
becomes complex and we have
\be m^2_\rho=\hat{m}^2_\rho+i\hat{m}_\rho\Gamma_\rho, 
\label{pole}
\ee 
where $\hat{m}_\rho$ (the ``mass" of the rho) and $\Gamma_\rho$ are real.

The $\pi^+\pi^-$ final state has isospin 1, so in order that G parity  be
conserved the $\rho$ meson must have isospin 1.  The photon can also couple to
the three pion system (as the electromagnetic current has isospin 1 and isospin
0 components). Once again, a similar enhancement is seen in the reaction \epp
attributed to the isospin zero $\omega$ , though with a much narrower peak
\cite{Barkov2} (and a mass, $\hat{m}_\omega> \hat{m}_\rho$).

As more data was collected on \ep and the resolution of the plot of the
cross-section improved (see Fig.~\ref{cross}), the interference of the $\omega$
meson in the reaction \ep was observed \cite{orsay}. One was now faced with a G
symmetry violating interaction, $\omega\ra\pi^+\pi^-$. This could not be
explained by the EM process $\omega\ra \gamma\ra\rho \ra2\pi$ as this is far
too small to account for what is seen experimentally. Thus, this interference
needs to be incorporated into the VMD picture.  This can be done by writing the
$\rho$ and $\omega$  meson propagators in matrix form \cite{OPTW,SW} and
generating the mixing by dressing the bare (isospin pure) matrix elements. We
thus have the bare matrix, $D^0_{\mu\nu}= -g_{\mu\nu}D^0$ where 
\be
D^0=\left(\begin{array}{cc}
(q^2-m_\rho^2)^{-1} & 0\\
0 & (q^2-m_\omega^2)^{-1}
\end{array} \right).
\ee
We then dress this to form the full (or physical) propagator given by
$D_{\mu\nu}=D^0_{\mu\nu}+D^0_{\mu\alpha}\Pi^{\alpha\beta}D_{\beta\nu}$
where the polarisation function, $\Pi_{\mu\nu}=(g_{\mu\nu}
-{q_\mu q_\nu}/{q^2})\Pi$, has off-diagonal elements
$\Pi_{\rho\omega}$, which generate
the mixing between the isospin pure $\rho_I$ and $\omega_I$.

Thus we introduce the decay mode $\gamma\ra\omega_I\ra\rho_I\ra 2\pi$ to model
what is seen in experiment. However we should also
consider the intrinsic decay $\omega_I\ra2\pi$. An argument by Renard, 
though,
claimed this effect will be suppressed and hence can be ignored
\cite{review,Renard} (this will be discussed further below).  
The dressed propagator, $D(q^2)$, will then have
off-diagonal elements, but the propagator could be diagonalised by
transforming to the physical basis,
$
\rho=\rho_I-\epsilon\omega_I$ and
$\omega=\omega_I+\epsilon\rho_I$.
The isospin violating mixing angle, $\epsilon$, is given by \cite{review}
\be
\epsilon=\frac{\Pi_{\rho\omega}}{m_\omega^2-m^2_\rho}
\label{eps}
\ee
where we use the complex masses of Eq.~(\ref{pole}). A model
for the pion form-factor could now be written down
\be
F_\pi(q^2)=\frac{\hat{m}_\rho^2}{g_\rho(q^2-m_\rho^2)}
+\epsilon\frac{\hat{m}_\omega^2}{g_\omega(q^2-m_\omega^2)}
\label{form}
\ee
which produced a remarkably good fit to the experimental data from only a few
parameters. In the absence, though, of any theoretical model,
$\Pi_{\rho\omega}$ has to be fitted to the data \cite{OPTW2}.

The mixing of elementary particles due to symmetry breaking had been considered
by Coleman and Schnitzer for the vector case \cite{CS}. They discussed two
kinds of mixing, mass (or particle) mixing, which was constant and current (or
vector) mixing which was momentum dependent. The conclusion reached was that
although mass mixing is perfectly adequate for spinless particles, current
mixing is better for spin one particles, because it does not violate the
conservation of electric charge (and hence $\Pi_{\rho\omega}$ should be
proportional to $q^2$).  Although the specific example they addressed was
$\omega\!-\!\phi$ mixing, they mentioned that off-diagonal (see above) current
mixing was suitable for the study of \rw mixing, which, at the time, had been
examined by Glashow \cite{G}.  It was a number of years before this suggestion
was followed up \cite{SW}, prompted by the direct experimental evidence for \rw
mixing in \ep \cite{orsay}.  

Coon {\it et al.} \cite{CSM} studied \rw mixing in the one boson exchange model
of the short-distance nuclear force as a contributing mechanism for the
generation of the charge symmetry violation (CSV) seen experimentally
\cite{TS}. The resulting potential is proportional to $\Pi_{\rho\omega}$, and
as it turns out, the value of $\Pi_{\rho\omega}$ extracted in the measurement
of the pion form-factor (Eq.~(\ref{form})) has the right sign and magnitude to
produce a reasonable fit to the data.

Although Coon {\it et al.} realised that the mixing function would in general
be momentum dependent, it was claimed that its value at the $\rho$ or $\omega$ 
mass point was all that was needed.  Nevertheless, its extraction in the pion
form-factor is for timelike $q^2$, while the vector mesons in the boson
exchange model of the $NN$ force have spacelike momentum.  Therefore, any
momentum dependence could have significant implications for the standard
treatment of CSV using \rw mixing. This was first realised by Goldman {\it et
al.} \cite{GHT} who constructed a simple model in which $\Pi_{\rho\omega}$ is
generated by a quark loop. The amplitude for the mixing is given by the
difference between the $\bar{u}u$ loop and the $\bar{d}d$ loop. In the limit of
isospin invariance, ($m_u=m_d$) the mixing would vanish.  The prediction of a
{\em significant} momentum dependence for \rw mixing forced Goldman {\it et
al.} to conclude that it would strongly reduce the standard class III and IV
CSV NN potential.  So can \rw mixing be simply assumed to be independent of
momentum, and if not, what does this say about nuclear models? 

A lot of model calculations were performed following this \cite{models} which
all produced similar results. At a more formal level, it was shown that for any
model in which the vector mesons coupled to a conserved current, the mixing
must vanish at $q^2=0$ \cite{OPTW}, precisely the constraint on vector mixing
expected by Coleman and Schnitzer \cite{CS}. In light of this, alternative
mechanisms for nuclear CSV were proposed, involving isospin violation at the
meson-nucleon vertex \cite{nuc}, rather than in the propagator. As both the
vertex and propagator parts of the $NN$ interaction are off-shell, they are
dependent on the choice of interpolating field for the vector mesons. It was
thus argued that one could find fields so that the sum of vertex and propagator
contributions is equivalent to a configuration in which all CSV occurs
through a momentum independent mixed propagator \cite{CM}. This argument,
though, has been disputed on the grounds of unitarity and analyticity
\cite{kim3}

However, the success of $NN$ models in which the CSV is generated by a
fixed-valued \rw mixing, provided considerable incentive to argue against
momentum dependence.  Miller \cite{miller} considered the mixing of the photon
and the rho. Traditional VMD has a fixed coupling between the photon and the
rho, but if this coupling were also generated by the kind of momentum dependent
loop processes used in for \rw mixing \cite{models} then the photon-rho
coupling would be strongly momentum dependent, hence destroying the successful
VMD phenomenology. However, an equivalent {\em momentum dependent} version of
VMD exists (which we shall refer to as VMD1; the traditional version we shall
call VMD2 \cite{review}), which was described by Sakurai thirty years ago
\cite{Sak}. VMD1 differs from VMD2 by having a linear-in-$q^2$ photon-rho
coupling {\em and} a direct coupling of the photon to the hadronic field,
unlike in VMD2 where photon-hadron interactions take place exclusively through
a vector meson.  As an example, the pion form-factor was plotted using VMD1
\cite{OPTW2} and the results are indistinguishable from the usual VMD2.  Thus
Miller's worry could be addressed by comparing the loop models \cite{models}
not to VMD2, but to VMD1 \cite{OWBK}. This makes sense not only because the
photon-$\rho$ mixings will be momentum dependent, but also because if the photon
is now allowed to couple to quarks (say) to form the loop, then we would it
expect it to be able to couple to the quarks in hadrons, hence introducing a
direct photon-hadron coupling not found in VMD2, but appearing in VMD1.

We might like, though, to make some model-independent statement about \rw. This
is difficult because the underlying theory,
QCD, is presently inaccessible at the
relevant energies. In the past 20 years we have developed some model
independent treatments of low energy strong interactions, and two of these
have been used to look at \rw mixing. The first is the technique of QCD
sum-rules (QCDSR) \cite{SVZ}. The basic idea is that one examines two-point
functions of various hadronic currents, expanding them out in powers of
$1/q^2$. At high $q^2$ QCD can be treated perturbatively due to asymptotic
freedom, but cannot be handled in this manner for low $q^2$  (for example,
around the $\rho$ mass).  So to work with the current correlators at low $q^2$ we
have to appeal to phenomenology (importantly the resonances which are related
to the vacuum structure). In this sense QCD sum rules are a bit of an art,
because there is no set method for using them.

Interestingly, one of the first examples of the use of QCDSR by the original
authors was \rw mixing. The problem is set up by considering the two-point
function
\be
C^{\mu\nu}_{\rho\omega}(q)=i\int d^4xe^{iq\cdot x}\langle0|{\rm T}(
J_\rho^\mu(x)J_\omega^\nu(0))|0\rangle,
\label{corr}
\ee
where
$J_\rho^\mu=(\bar{u}\gamma_\mu u-\bar{d}\gamma_\mu d)/2$ and 
$J_\omega^\mu=(\bar{u}\gamma_\mu u+\bar{d}\gamma_\mu d)/6$.
This {\em current } correlator (Eq.~(\ref{corr})) was then used by Hatsuda {\it
et al.} \cite{HHMK} to examine the momentum dependence of \rw mixing by
equating it with the mixed propagator (after extracting the transverse
tensor $(g_{\mu\nu}-q_\mu q_\nu/q^2)$)
\be
D_{\rho\omega}(q^2)=\frac{\Pi_{\rho\omega}
(q^2)}{(q^2-\hat{m}_\rho^2)(q^2-\hat{m}_\omega^2)}.
\label{diag}\ee
As pointed out by Maltman \cite{kim1}
though, the association of the correlator with the off-diagonal propagator is
only relevant if one uses interpolating fields for the $\rho$ and $\omega$  mesons
proportional to the currents $J_\rho^\mu$ and $J_\omega^\mu$, otherwise the
correlator cannot be used to provide information about the off-shell behaviour
of the mixing element of the vector meson propagator.  Hatsuda {\it et al.} 
concentrated on the effect of the $\rho$ and
$\omega$ in this correlator, which as the most nearby resonances might be expected to play the
dominant role. However, Maltman found that the $\rho$ and $\omega$  contributions
actually partially cancel. Because of this, the $\phi$, although quite far
away and hence contributing with a much lesser strength than the individual $\rho$ and $\omega$
becomes important for the isospin-breaking correlator (an
effect not considered in the previous two analyses \cite{SVZ,HHMK}).  

In all analyses the sum rule result is ultimately compared to the data for the
G-parity violation seen in \ep. The correlator, though, is only relevant to the
contribution from the mixing of the isospin pure states, $\rho_I\!-\!
\omega_I$, to the isospin breaking seen in the process. The competing process,
$\omega_I \ra \pi^+\pi^-$, is overlooked (as mentioned earlier), but  Maltman
found the $\rho_I\!-\! \omega_I$ contribution (as determined by the current
correlator in QDCSR) under-estimates the isospin-violation seen
experimentally. We shall discuss the matter of intrinsic decay further below.

Leinweber {\it et al.} in two recent papers \cite{IJL} examined the effects of
including the widths of the $\rho$ and $\omega$  mesons in the QCDSR
calculation performed by Hatsuda {\it et al.}. They replaced the real parts of
the mass in Eq.~(\ref{diag}), by the complex pole positions given in
Eq.~(\ref{pole}). Following Maltman \cite{kim1}, they included the $\phi$
mesons, but found that its contribution was negligible. Perhaps of most
interest, following the nuclear CSV debate, was their claim that for certain
values of $\lambda$, $\Pi_{\rho\omega}$ has the same sign and similar magnitude
in the space-like region to the on-shell value.

Because the reaction \ep is the only place \rw is actually seen, it was decided
that a new {\em and general} analysis should be performed \cite{MOW}. The two
effects normally ignored, momentum dependence of $\Pi_{\rho\omega}$ and the
intrinsic decay $\omega_I \ra \pi^+\pi^-$ were included. As two recent fits to
data had been performed \cite{fits} all that needed to be done was to construct
a precise theoretical expression for the form-factor, which could be compared
to the numbers extracted from these analyses for the expression
\be
F_\pi(q^2)\propto P_\rho+Ae^{i\phi}P_\omega,
\label{formf}
\ee
where $P_{\rho,\omega}$ are the poles of the $\rho$ and $\omega$  propagators.  The
starting point was  the mixed matrix formalism. In the data one sees two
resonance peaks - a broad one associated with the physical (as opposed to
isospin-pure) rho, and a narrow one associated with the physical $\omega$. 
Thus,
the mixed propagator, should only contain two poles in the physical bases. We
choose this basis to be $\rho=\rho_I -\epsilon_1 \omega_I$ and $\omega=
\omega_I+ \epsilon_2 \rho_I$ and write the propagator in the new, physical
basis. This allows us to fix $\epsilon_{1,2}$ by demanding that there are no
poles in the off-diagonal pieces, i.e. that all resonant behaviour is
associated with the physical mesons. This gives expressions for
$\epsilon_{1,2}$ similar to Eq.~(\ref{eps}) but with arguments for
$\Pi_{\rho\omega}$ at $m_\omega^2$ and $m_\rho^2$ respectively,
as noted by Harte and Sachs \cite{HS}. Thus, in
general the bases are not related by a simple rotation, and the pure and real
bases are not equivalent (as the transformation 
between them is not orthogonal).  For the
case that $\Pi_{\rho\omega}$ is either fixed or linear in $q^2$, the
off-diagonal terms can be made to disappear completely, but in general they
survive and contribute to the non-resonant (i.e. no singular piece) background
(although this is only a minor effect). The second part of the analysis centres
on the decay $\omega_I \ra \pi^+\pi^-$. A closer examination of the Renard
argument shows that the cancellation is not exact, and a reasonable fraction of
the intrinsic decay survives, which adds to the total interaction and this
turns out to be {\em crucial} to the analysis. 
In a world of exact experimental precision the pre-factor, $Ae^{i\phi}$, of
the $\omega$  pole in the expression for the form-factor (Eq.~(\ref{formf})),
would enable us to pin down the values of the two unknowns, $\Pi_{\rho\omega}$
and the strength of $\omega_I \ra \pi^+\pi^-$. Unfortunately this is not the
case, and the considerable uncertainty in the Orsay phase, $\phi$, and the
lesser uncertainty in $A$ allows a whole spread of values for the two unknowns;
$\Pi_{\rho\omega}$ can take values in the set $(-840,-6240)$ MeV$^2$. 
Naturally if there were no contribution from $\omega_I \ra \pi^+\pi^-$ we would
recover the usual analysis and obtain $\Pi_{\rho \omega}=-3960$ MeV$^2$ (c.f.
the value -4520$\pm$600 MeV$^2$ \cite{CB}). In light of this Maltman's QCDSR
analysis is quite interesting, as it seems to provide theoretical evidence for
a non-zero contribution from intrinsic decay, leading us to re-think the
present status of the traditional extraction of $\Pi_{\rho\omega}$.  It also
brings in to question the value of $\Pi_{\rho \omega}$ used 
in nuclear models.

Another model-independent method for treating the strong interaction at low
energies is Chiral Perturbation Theory (ChPT). It is the subject of many recent
reviews \cite{chiral}, and essentially it sets up an effective model involving
the pseudo-scalar octet and admitting all terms allowed by the symmetry of the
original QCD Lagrangian, organised as a perturbative series in $q^2$. However,
the symmetries of QCD in question (chiral symmetry, isospin symmetry) are not
exact; the main feature of ChPT is that it breaks these symmetries, for a meson
theory, exactly as QCD breaks them. The various free parameters of the theory
are then fixed by comparison to experiment.

As a perturbative series in $q^2$, ChPT is only reliable in the low momentum
region. The relatively heavy vector mesons, therefore, do not fit naturally
into it. As resonances of QCD processes (which is really what they represent),
they play a very important role in strong physics, but unfortunately ChPT
breaks down well before the $q^2$ of the poles we associate with vector mesons.
Thus, it is usually with an assumption such as VMD that the vectors mesons are
fitted into ChPT, to create an effective model incorporating ChPT. One such
model, in which the vector mesons appear in the Lagragian as antisymmetric
tensors has been used by Urech to study \rw mixing \cite{urech}.

The hadronic currents, as appearing in the correlator, have no such difficulty
and appear quite naturally in ChPT. Maltman \cite{kim1} thus used ChPT as a
complement to his QCDSR calculation of the mixed correlator. He examined the
correlator to one-loop order ($O(q^4)$) in ChPT obtaining a result dependent on
both $q^2$ and the mass difference of the neutral and charged kaon (vanishing
when these masses are equal, which, in ChPT occurs when $m_u=m_d$). The result
was much smaller than the corresponding QCDSR result, indicating that the next
order contribution (two-loops, $O(q^6)$) needed to be included as well. The
two-loop calculation \cite{kim2} seems to show that the chiral series is not
convergent enough to allow one to truncate even at $O(q^6)$ (which is the
present-day limit of ChPT).

The study of \rw mixing has pushed VMD and available data to its limit. So far
we have established, within the matrix approach to VMD, that we expect the
mixing of the pure isospin states to satisfy $\Pi_{\rho\omega}(0)=0$ (for
models in which the mesons couple to conserved currents), and that, due to
experimental uncertainty in the pion form-factor, we cannot distinguish between
$\rho_I-\omega_I$ mixing and the intrinsic decay $\omega_I \ra\pi^+\pi^-$. As
we have seen, QCD sum rules and ChPT are providing new insights to this old
problem, but their results are open to interpretation. Physical processes in
total, rather than their individual contributions are what we actually observe
and so our studies need to reflect this.  It remains for a completely
consistent field theory based description of isospin violation in both the
timelike (\rw mixing) and spacelike (nuclear CSV) to be constructed. We
look forward to future developments in this field.

\vspace{1cm}

{\bf Acknowledgements} I would like to thank A.W.~Thomas, A.G.~Williams
(Adelaide U.), C.D.~Roberts (Argonne) and K.R.~Maltman (York U., Canada) for
helpful conversations. This work is supported in part by the Australian
Research Council.

\end{document}